# INCOMPLETE HIPPOCAMPAL INVERSION AND HIPPOCAMPAL SUBFIELD VOLUMES: IMPLEMENTATION AND INTER-RELIABILITY OF AUTOMATIC SEGMENTATION

*Fragueiro, Agustina[1]; Committeri, Giorgia[2]; Cury, Claire[1]*

1. Univ Rennes, CNRS, Inria, Inserm, IRISA UMR 6074, Empenn - ERL U 1228, F-35000 Rennes, France.
2. Department of Neuroscience, Imaging and Clinical Sciences, University G. d'Annunzio of Chieti-Pescara, Italy.

## ABSTRACT

The incomplete hippocampal inversion (IHI) is an atypical anatomical pattern of the hippocampus. However, the hippocampus is not a homogeneous structure, as it consists of segregated subfields with specific characteristics. While IHI is not related to whole hippocampal volume, higher IHI scores have been associated to smaller CA1 in aging. Although the segmentation of hippocampal subfields is challenging due to their small size, there are algorithms allowing their automatic segmentation. By using a Human Connectome Project dataset of healthy young adults, we first tested the inter-reliability of two methods for automatic segmentation of hippocampal subfields, and secondly, we explored the relationship between IHI and subfield volumes. Results evidenced strong correlations between volumes obtained thorough both segmentation methods. Furthermore, higher IHI scores were associated to bigger subiculum and smaller CA1 volumes. Here, we provide new insights regarding IHI subfields volumetry, and we offer support for automatic segmentation inter-method reliability.

## 1. INTRODUCTION

The incomplete hippocampal inversion (IHI) is an atypical anatomical pattern of the hippocampus. Specifically, it is a developmental anomaly characterized by the incomplete infolding of the hippocampal subfields, which gives the hippocampus a more round or pyramidal shape [1]. IHI is characterized by medial positioning of the hippocampus and a deep collateral sulcus, and occurs predominantly in the left hippocampus [2]. Although it is considered the end of the normal phenotypic spectrum, it is more prevalent in epilepsy and it is thought to be a factor of susceptibility for hippocampal sclerosis [3]. Furthermore, it has been associated with hippocampal loss of volume in aging. Specifically, using manually segmentation of hippocampal subfields on healthy elder adults, it has been reported a relationship between IHI scores and bilateral CA1 volumes [4], a subfield that demonstrates notable loss of volume in normal aging [5]. On the contrary, whole hippocampal volume was not related to IHI scores [4].

The hippocampus consists of distinct and functionally segregated subfields with specific cell properties [6]. Currently, there are algorithms allowing the automatic segmentation of hippocampal subfields and related medial temporal lobe subregions. However, this segmentation is challenging due to their small size and lack of contrast for delineation due to magnetic resonance imaging (MRI) signal loss in medial temporal regions [7]. Automatic Segmentation of Hippocampal Subfields (ASHS) method, proposed by the University of Pennsylvania[8], requires T1w and T2w images, and has been validated in a population of healthy elderly controls and in clinical patients with prodromal and dementia Alzheimer's Disease [9]. Another widely used automatic segmentation method is proposed by FreeSurfer [10], which offers hippocampal subfields delineation with multiple input options. Reliability of volumetric estimations has been reported for FreeSurfer's procedure using images acquired by two different MR scanners [7], as well as using different input images (T1w, T2w, T1w and T2w)[11].

By using a dataset collected in a healthy young adult population by the Human Connectome Project (HCP)[12], we tested the inter-method (i.e. ASHS and FreeSurfer) reliability for volumetric analysis using automatic segmentation of hippocampal subfields. In a second moment, we explored the relationship between IHI scores, rated according to the criteria reported in Cury et al. (2015), and hippocampal subfields' volumes extracted with both ASHS and FreeSurfer methods.

## 2. MATERIALS AND METHODS

### 2.1. Participants

Preprocessed 3T MRI scans belonging to a total of 390 healthy young adults (age=26-30, 217 females) have been downloaded from the Human Connectome Dataset WU-Minn [12].

### 2.2. MRI data

All HCP subjects were scanned on a customized Siemens 3T housed at Washington University in St. Louis using the gradients of the WU-Minn and MGH-UCLA Connectome scanners. T1w 3D MPRAGE were acquired with TR=2400 ms, TE=2.14 ms, TI=1000 ms, flip angle=8 deg, FOV=224x224, 0.7 mm isotropic voxel, bandwidth=210 Hz/px, iPAT=2, and acquisition time=7:40 (min:sec). T2w 3D T2-SPACE were acquired with TR=3200 ms, TE=565 ms, variable flip angle, FOV=224x224, 0.7 mm isotropic voxel, bandwidth=744 Hz/px, iPAT=2, and acquisition time=8:24(min:sec).

Preprocessing of data was done using Version 3.1-3.21 of the minimal preprocessing pipelines including spatial artifact/distortion removal, surface generation, cross-modal registration, and alignment to standard space, detailed in Glasser et al. (2013) [13]. T1w and T2w in their native space and original dimensions after rigid-body rotation to AC-PC alignment were used for segmentation.

### 2.3. Hippocampal Subfields Segmentation

*2.3.1. Automatic Segmentation of Hippocampal Subfields*
The open-source ASHS software, developed by the Penn Image Computing and Science Laboratory (PICSL) at the University of Pennsylvania [8], invoke image analysis algorithms from FSL [14] and ANTS [15]. ASHS implements the joint label fusion (JLF) [16] and corrective learning (CL) [17] algorithms. The steps of the ASHS segmentation pipeline are described in Yushkevich et al. (2015). Preprocessed T1w and T2w in AC-PC line were provided as input for the segmentation. ASHS was run using the IKND Magdeburg Young Adult 7T Atlas, provided by Ottovon-Guericke-University Magdeburg. The atlas has been developed by manual segmentation of hippocampal and the adjacent medial temporal lobe regions [18]. The subregions segmented with this atlas are: entorhinal cortex (ErC), area 35 and 36 of the perirhinal cortex (PrC), parahippocampal cortex (PhC), subiculum (Sub), cornu ammonis (CA) 1, 2 and 3, dentate gyrus (DG), cysts, and tail. Volumes in $mm^3$ for each subfield were directly provided by ASHS, and JLF-CL *corr_nogray* output were used for the following analysis as data appearance with the training set was not assumed.

*2.3.2. FreeSurfer segmentation of hippocampal subfields*
The tool for segmentation of hippocampal subfields and nuclei of the amygdala offered by FreeSurfer 7, was run using the T1w scan from recon-all method (*segmentHA_T1.sh*). Preprocessed T1w in AC-PC line were provided as input. This tool segments the different subfields by using a Bayesian inference approach based on image intensities and prior knowledge of a probabilistic atlas which was generated from in vivo manual segmentations and ultra-high-resolution ex vivo MRI data [19]. The subfields segmented with this method followed the head-body-tail subdivision [10] and are: CA1 body, CA1 head, CA2/3 head, CA2/3 body, CA4 head, CA4 body, fimbria, granule cell layer of dentate gyrus (GC-ML-DG) head, GC-ML-DG body, hippocampus-amygdala-transition-area (HATA), tail, hippocampal fissure, molecular layer head, molecular layer body, parasubiculum, presubiculum head, presubiculum body, subiculum head, subiculum body. Volumes in $mm^3$ for each hippocampal subfield were obtained through FreeSurfer (*quantifyHAsubregions.sh*).

### 2.4. Incomplete Hippocampal Inversion Rating

T1w images in AC-PC were linearly transformed to MNI using *flirt* in FSL. Using medInria for visualization, IHI was manually rated according to the criteria reported in Cury et al. (2015). Specifically, criteria C1, C2, C3, and C5 were evaluated independently. C1 assesses the roundness and verticality of the hippocampal body, C2 assesses the verticality and depth of the collateral sulcus relatively to the size of the hippocampus, C3 assesses the medial positioning of the hippocampus, and C5 considers the depth of the collateral and the occipitotemporal sulcus with respect of the subiculum. C4 was not rated. A total IHI score (from 0 to 8) was calculated for each hemisphere by summing all criteria. If the total score exceeded 3.75, the hemisphere was classified as IHI [2].

Before starting the rating of the database, we evaluated intra- and inter-rater agreement. For intra-rater coherence, A.F. rated the same database of 30 subjects with a distance of 15 days. For inter-rater agreement, C.C. rated the same 30 subjects. Kappa scores were calculated for C5, and kappa weight scores were calculated for criteria C1, C2, and C3. All intra- and inter-rater scores were assessed independently for each hemisphere and all of them were above 0.65.

### 2.5. Statistical Analysis

*2.5.1. ASHS and FreeSurfer: inter-method reliability*
The following analysis were conducted on 381 subjects as segmentation failed in 7 subjects, and 2 more subjects were excluded after visual assessment of segmentation quality. As subfield definition differed among methods, we first combined the subfields by summing their volumes to obtain four common subfields (Table 1).

**Table 1.** Re-definition of segmented subfields to get comparable subfield volumes for both ASHS and Freesurfer methods.

| Common subfields | ASHS | Freesurfer |
|---|---|---|
| *CA1* | CA1 | CA1 head + body |
| *CA2/3* | CA2+CA3 | CA2/3 head + body |
| *Subiculum* | Subiculum | Subiculum head +body |
|  |  | Presubiculum head +body |
|  |  | Parasubiculum |
| *Tail* | Tail | Tail |

Normal distribution of data was evaluated with the Kolmogorov-Smirnov Test. As not all subfield volumes were normally distributed, we conducted non-parametric statistics. Spearman's correlations between segmentation methods (ASHS vs. Freesurfer) were conducted for all four common subfields (CA1, CA2/3, subiculum, tail) and for whole hippocampal volumes (CA 1, 2, 3 and 4, DG, complete subiculum, tail). In addition, Mann-Whitney tests were conducted for each shared subfield and for the whole hippocampus comparing volumes obtained through both methods.

*2.5.2. IHI and hippocampal subfields' volumes*
First, we conducted Mann-Whitney tests for independent samples between IHI and no-IHI hippocampus, independently for each hemisphere, comparing volumes of the four shared subfields (CA1, CA2/3, subiculum, tail) and the whole hippocampus. Tests were conducted separately for volumes extracted with each segmentation method (i.e. ASHS, FreeSurfer).

Finally, four linear regression models were tested: 1) IHI scores for the left hemisphere were included as dependent continuous variable and subfields of the same hemisphere segmented with ASHS as independent variables (ErC, A35, A36, PhC, subiculum, CA1, CA2, CA3, DG, cysts, tail); 2) IHI scores of the right hemisphere as dependent variable and volumes corresponding to the same hemisphere extracted with ASHS as independent variables; 3) IHI scores of the left hemisphere as dependent variable and volumes corresponding to the same hemisphere obtained through FreeSurfer as independent variables (CA1_body, CA1 head, CA2/3 head, CA2/3 body, CA4 head, CA4 body, fimbria, GC-ML-DG head, GC-ML-DG body, HATA, tail, hippocampal fissure, molecular layer head, molecular layer body, parasubiculum, presubiculum head, presubiculum body, subiculum head, subiculum body); 4) IHI scores of the right hemisphere as dependent variable and volumes extracted with FreeSurfer as independent variables. Volumes were previously corrected by whole brain volume (BV) calculated on FSL using *fslstats* function. Gender was also included as independent variable in all models, but age was not included due to the very narrow range. The forward stepwise method was used to identify the best grouping of independent variables that account for the most variance in the outcome (R-squared). This approach enters independent variables one at a time after considering the marginal contribution of a variable controlling for other variables already in the model. The absence of multicollinearity between variables included in each model was tested though the variance inflation factor (all VIF<5.00). Finally, using the automatic linear modeling of IBM SPSS Statistics 25, the predictor importance of each independent variable was determined by computing the reduction in variance of the target attributable to each independent variable, via a sensitivity analysis [20]. The score indicates the relative importance of each variable in estimating the model, with values for all predictors on the display summing 1.

All p-values reported were Bonferroni corrected for multiple comparisons (10 correlations + 24 Mann-Whitney tests + 2 hemispheres x 11 subfields segmented with ASHS + 2 hemispheres x 19 subfields segmented with Freesurfer = total of 94). All statistical analyses were conducted on IBM SPSS Statistics 25.

## 3. RESULTS

### 3.1. ASHS vs. Freesurfer
Spearman's correlations were significant for all common subfields and for the whole hippocampus (Table 2).

Mann-Whitney tests evidenced significant differences between volumes extracted through both segmentation methods (i.e. ASHS vs. FreeSurfer) for all subfields (all p<0.001) (Figure 1). Depending on the subfield, one method segmented bigger or smaller area with respect to the other (Figure 2). Furthermore, whole hippocampal volumes obtained thought FreeSurfer segmentation were significantly bigger for both hemispheres (Left: U=22332, p<0.001, Right: U=17708, p<0.001).

**Table 2.** Spearman's correlations between volumes obtained through ASHS and FreeSurfer segmentation methods.

| Hemisphere | Subfields | Spearman's rho | p-values |
|---|---|---|---|
| **Left** | *CA1* | .82 | <0.001 |
| | *CA2/3* | .63 | <0.001 |
| | *Subiculum* | .79 | <0.001 |
| | *Tail* | .78 | <0.001 |
| | *Whole hipp.* | .91 | <0.001 |
| **Right** | *CA1* | .87 | <0.001 |
| | *CA2/3* | .67 | <0.001 |
| | *Subiculum* | .80 | <0.001 |
| | *Tail* | .76 | <0.001 |
| | *Whole hipp.* | .92 | <0.001 |

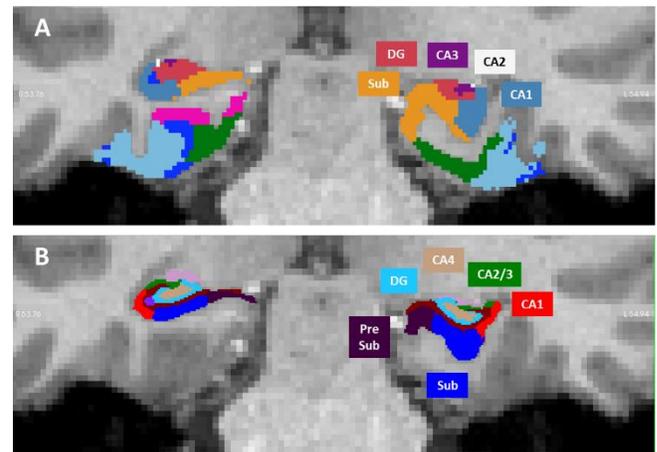

**Figure 1.** Coronal view of the segmentations obtained with: A) ASHS, and B) FreeSurfer at the hippocampal body level. The left hippocampus was classified as IHI while the right as no IHI.

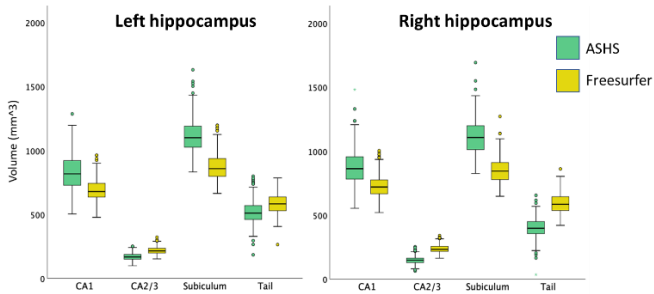

**Figure 2.** Boxplot comparing subfields' volumes segmented with ASHS vs. Freesurfer.

### 3.2. IHI and hippocampal subfields

The 23% of left hippocampi and 9% of the right hippocampi were rated as IHI. A 6% of brains were identified as bilateral IHI. The rates of IHI prevalence in the current sample are similar to those previously reported in the healthy population[2], [4].

Mann-Whitney tests evidenced that subiculum volumes extracted with a FreeSurfer were significantly higher for IHI comparing to not IHI hippocampi bilaterally (Left: U=8442, p<0.001; Right: FreeSurfer, U=3112, p<0.001) (Figure 3). CA2/3 volumes extracted with ASHS were smaller in IHI hippocampi for in the right hemisphere (U=3898.5, p=0.048). No significant differences were found for the other subfields (i.e. CA1, tail), neither for whole hippocampus volume.

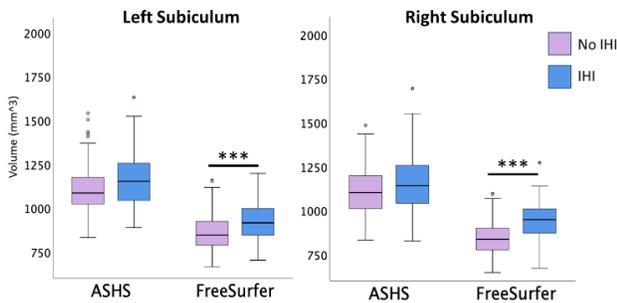

**Figure 3.** Boxplot comparing subiculum volumes between IHI and no IHI hippocampus for both ASHS and FreeSurfer methods.

The two regression models including IHI scores as dependent variable and subfield volumes obtained with ASHS as independent variables, were significant bilaterally (Left: $R^2$=0.23, p=0.005; Right: $R^2$=0.168, p=0.025). Specifically, the stepwise method identified that for both hemispheres smaller CA2 volumes were associated to higher IHI scores (Left: β= -0.398, p<0.001, importance=0.648; Right: β= -0.26, p<0.001, importance=0.42). For the right hemisphere, smaller CA1 volumes were also significantly associated to IHI severity (Left: β= -0.174, p=0.094, importance=0.155; Right: β= -0.254, p<0.001, importance=0.385).

As well, the two regression models conducted using volumes extracted with FreeSurfer were significant for both left ($R^2$=0.537, p=0.014) and right ($R^2$=0.508, p=0.027) hemispheres. Specifically, the stepwise method identified that smaller CA1 (Left: β= -0.477, p<0.001, importance=0.308; Right: β= -0.435, p<0.001, importance=0.341) and bigger subiculum (Left: β=0.441, p<0.001, importance=0.361; Right: β=0.398, p<0.001, importance=0.415) volumes, both of them exclusively at the level of the hippocampal body, were associated to higher IHI scores bilaterally (Figure 4). In addition, the model conducted for the left hemisphere identified, but with lower importance score (<0.05), the tail (β=-0.189, p<0.006) and the subiculum at the head level (β=-0.155, p=0.008). The model conducted for the right hemisphere identified also the fimbria, but with lower importance (β=0.199, p=0.001, importance=0.115).

## 4. DISCUSSION

Although volumes extracted from common subfields differed among methods, they highly correlated. We understand that this difference may be related to the different atlas used by the two methods. However, the inter-reliability among methods is supported by the high correlation scores for all common subfields and the whole hippocampus.

Regarding the relationship between subfield volumes and IHI, we found bigger subiculum volumes for IHI comparing to no IHI hippocampi bilaterally. Furthermore, the regression models including IHI scores as a continuum, identified that lower CA1 and higher subiculum volumes were associated with higher IHI scores bilaterally. Interestingly, Freesurfer's head-body delineation allowed us to further specify this relationship exclusively for the hippocampal body, portion of the hippocampus where IHI is rated. An association between higher IHI scores and smaller CA1 has already been described by implementing manual segmentation in healthy elder adults [4]. Our results extended these findings to a population of young adults. In addition, we observed an association between bigger subiculum volumes and IHI severity. This result is consonant with the incomplete development of IHI hippocampi before gestational week 21[1], [21], when the subiculum is larger, and CA1, CA2, and CA3 are arranged linearly. The incompletion of subfields infolding, results in IHI characteristic fetal shape.

Interestingly, we observed that smaller CA2 volumes were associated to higher IHI scores bilaterally. This result was only evidenced using ASHS method, which allows CA2 segmentation, independently from CA3. Considering the challenge of CA2 delineation due to its small size, further analyses are needed.

Structural imaging is a potentially important biomarker in clinical practice, and automatic segmentation methods represent a tool that may boost its potential. Here, we provided new insights regarding the relationship among IHI and hippocampal subfield volumes, and in addition, we offer support for inter-reliability of automatic segmentation methods. Future research will explore the relationship between these hippocampal features and related cognitive functions, specifically memory and spatial navigation abilities.

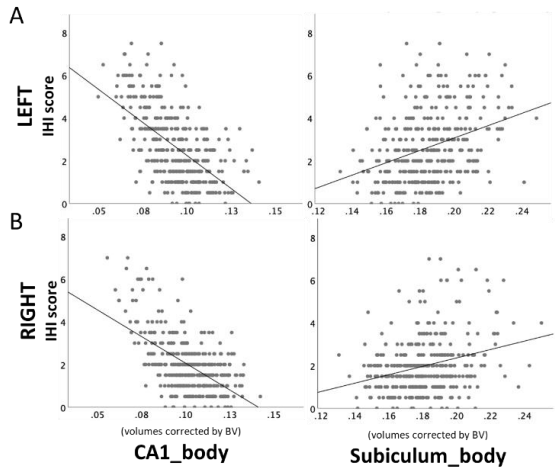

**Figure 4.** Scatterplot of CA1 and subiculum volumes at the level of the hippocampal body, extracted with FreeSurfer, in relationship to IHI scores for A) left and B) right hemispheres.

## 5. COMPLIANCE WITH ETHICAL STANDARDS

This study was conducted retrospectively using human subject data made available in open access by the Human Connectome Dataset WU-Minn [12]. Ethical approval was not required as confirmed by the license attached with the open access data.

## 6. ACKNOWLEDGMENTS

No funding was received for conducting this study. The authors have no relevant financial or non-financial interests to disclose.